# A High-Entropy Silicide: $(Mo_{0.2}Nb_{0.2}Ta_{0.2}Ti_{0.2}W_{0.2})Si_2$

Joshua Gild[a], Jeffrey Braun[c], Kevin Kaufmann[b], Eduardo Marin[b], Tyler Harrington[a], Patrick Hopkins[c], Kenneth Vecchio[a, b], Jian Luo[a, b, *]

[a]Program of Materials Science and Engineering, University of California, San Diego, La Jolla, CA 92093-0418, USA;

[b]Department of NanoEngineering, University of California, San Diego, La Jolla, CA 92093-0448, USA;

[c]Department of Mechanical and Aerospace Engineering, University of Virginia, Charlottesville, VA 22904, USA;

A high-entropy metal disilicide, $(Mo_{0.2}Nb_{0.2}Ta_{0.2}Ti_{0.2}W_{0.2})Si_2$, has been successfully synthesized. X-ray diffraction (XRD), energy dispersive X-ray spectroscopy (EDX), and electron backscatter diffraction (EBSD) collectively show the formation of a single high-entropy silicide phase. This high-entropy $(Mo_{0.2}Nb_{0.2}Ta_{0.2}Ti_{0.2}W_{0.2})Si_2$ possesses a hexagonal C40 crystal structure with ABC stacking sequence and a point group of $P6_222$. This discovery expands the known families of high-entropy materials from metals, oxides, borides, carbides, and nitrides to a silicide, for the first time to our knowledge, as well as demonstrating that a new, non-cubic, crystal structure (with lower symmetry) can be made into high-entropy. This $(Mo_{0.2}Nb_{0.2}Ta_{0.2}Ti_{0.2}W_{0.2})Si_2$ exhibits high nanohardness of $16.7 \pm 1.9$ GPa and Vickers hardness of $11.6 \pm 0.5$ GPa. Moreover, it has a low thermal conductivity of $6.9 \pm 1.1$ W m$^{-1}$ K$^{-1}$, which is approximately one order of magnitude lower than that of the widely-used tetragonal $MoSi_2$ and ~1/3 of those reported values for the hexagonal $NbSi_2$ and $TaSi_2$ with the same crystal structure.

[*]Corresponding author; E-mail address: jluo@alum.mit.edu (J. Luo).
A preprint archived at: arXiv:1902.01033

# Graphical Abstract

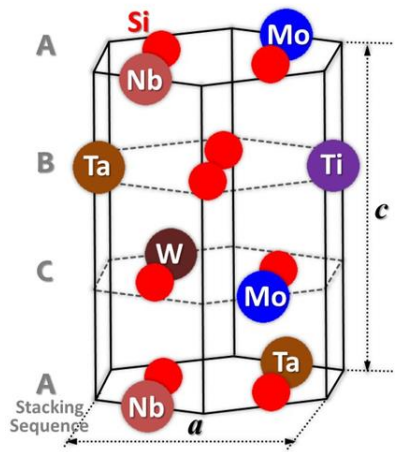
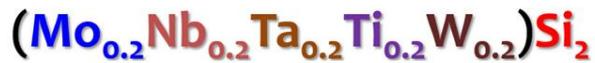
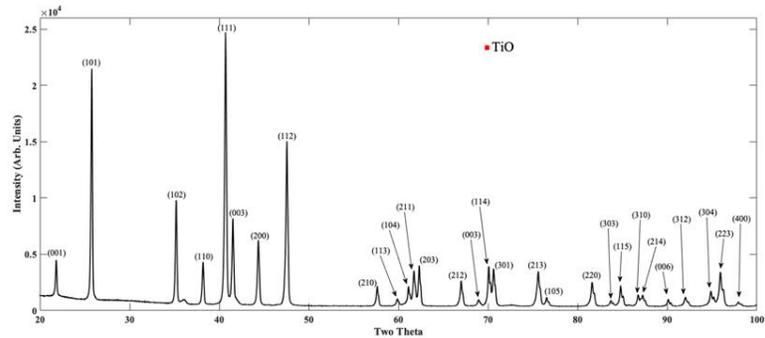

**Highlights**

- A high-entropy silicide, (Mo0.2Nb0.2Ta0.2Ti0.2W0.2)Si2, synthesized
- Expanding high-entropy materials to a silicide for the first time
- A new, non-cubic, high-entropy crystal structure demonstrated
- High nanohardness of 16.7 ± 1.9 GPa and Vickers hardness of 11.6 ± 0.5 GPa
- Much-reduced thermal conductivity of 6.9 ± 1.1 W m-1 K-1 (~1/10 of that of MoSi2)

## 1. Introduction

Research on high-entropy alloys (HEAs), also known as multiple principal element alloys (MPEAs) or complex concentrated alloys (CCAs), has attracted considerable interest in the last ~15 years due to their unique properties and large compositional space for engineering [1-8]. A majority of the metallic HEAs adopt the simple face-centered cubic (FCC) or body-centered cubic (BCC) crystal structures, and a few hexagonal close packed (HCP) HEAs have also been made [1-8].

Only in the last ~3.5 years have the ceramic counterparts to the metallic HEAs, or "high-entropy ceramics," been successfully fabricated in bulk forms. In 2015, Rost *et al.* reported an entropy-stabilized oxide, $(Mg_{0.2}Ni_{0.2}Co_{0.2}Cu_{0.2}Zn_{0.2})O$, of a rocksalt structure (with a FCC Bravais lattice) [9]. In 2016, high-entropy metal diborides, *e.g.* $(Hf_{0.2}Zr_{0.2}Ta_{0.2}Nb_{0.2}Ti_{0.2})B_2$, were reported as a new class of ultra-high temperature ceramics (UHTCs) and the first high-entropy borides (as well as the first non-oxide high-entropy ceramics made in the bulk form) [10]. Subsequently, the research on high-entropy ceramics has made rapid progresses and attracted increasing attention. First, the high-entropy (entropy-stabilized) rocksalt oxides have been studied extensively due to their great potentials as functional materials with low thermal conductivities [11-13] and colossal dielectric constants [14], as well as their potential applications in lithium-ion batteries [15,16]. Second, high-entropy metal diborides have also been studied by many groups as a new class of promising structural ceramics with increased hardness [17-19]; this line of work has further stimulated the subsequent development of high-entropy metal carbides as another class of UHTCs with increased hardness by various groups worldwide [20-29]. Third, several other classes of high-entropy ceramics have also been reported, including perovskite [30-32], spinel [33], defective fluorite-structured [34,35], and rare earth [32,36] oxides, as well as high-entropy nitrides [37,38]. It is worth noting that the high-entropy oxides [30-32,34-36], carbides [20-29], and nitrides [37,38] discovered to date all have cubic crystal structures with high symmetries. The only exception is the high-entropy metal diborides, which have a hexagonal ($AlB_2$) crystal structure, yet with a rather high symmetry (P6/mmm) [9].

As an increasing number of high-entropy oxides [30-32,34-36], borides [10,17-19], carbides [20-29], and nitrides [37,38] have been discovered, this study first reports the synthesis and characterization of a high-entropy silicide: $(Mo_{0.2}Nb_{0.2}Ta_{0.2}Ti_{0.2}W_{0.2})Si_2$, for the first time to our knowledge. Moreover, this $(Mo_{0.2}Nb_{0.2}Ta_{0.2}Ti_{0.2}W_{0.2})Si_2$ possesses a $CrSi_2$-type hexagonal C40 structure with the ABC stacking sequence (Fig. 1); it represents a more complex crystal structure (with a lower $P6_222$ symmetry) in comparison with those reported in prior studies, thereby extending the state of the art for the discovery of new high-entropy materials.

In general, refractory disilicides, particularly $MoSi_2$, are of great interest for high-temperature applications [39-45]. In this study, we have also examined the properties of this new high-entropy $(Mo_{0.2}Nb_{0.2}Ta_{0.2}Ti_{0.2}W_{0.2})Si_2$, showing high hardness (16.7 ± 1.9 GPa nanohardness and 11.6 ± 0.5 GPa Vickers hardness) and much reduced thermal conductivity (6.9 ± 1.1 W m$^{-1}$ K$^{-1}$).

## 2. Experimental Procedure

Powders of $MoSi_2$, $NbSi_2$, $TaSi_2$, $TiSi_2$, $WSi_2$, and $ZrSi_2$ (99% purity, ≥45 μm; Alfa Aesar) were utilized as starting materials. The raw powders were mixed via high-energy ball milling (HEBM) utilizing a SPEX 8000D mill for 6 hr in a silicon nitride jar with silicon nitride media. Heptane was used to create a slurry for grinding to prevent caking of the powders and to minimize oxidation in the milling containers. The HEBM was done in 30-minute intervals, interrupted by 10-minute resting pauses to avoid overheating. The powders were then densified into 20-mm diameter disks via spark plasma sintering (SPS, Thermal Technologies, CA, USA) at 1650 °C for 10 min under a uniaxial pressure of 50 MPa; then the pressure was immediately reduced to 10 MPa at a rate of 40 MPa/min at 1650 °C to minimize creep. The chamber was initially pumped down to vacuum of at least 20 mTorr and backfilled with argon for three times prior to the SPS experiments to minimize oxidation and a vacuum was maintained throughout the sintering process. The graphite die was lined with 125 μm thick graphite paper to prevent reaction of the specimen with the die.

The silicide was characterized by X-ray diffraction (XRD) utilizing a Rigaku diffractometer with Cu Kα radiation. Scanning electron microscopy (SEM) was carried out, and the corresponding energy dispersive X-ray (EDX) spectroscopy compositional maps and electron backscatter diffraction (EBSD) maps were collected. The EDX measurements were performed at an e-beam voltage of 20 kV to examine the higher energy peaks of Hf, Ta, and W for minimal convolution of the peaks.

Hardness and modulus measurements were conducted via nano-indentation on a KLA-tencor G200 Nanoindenter (KLA-tencor, CA, USA). Nanohardness measurements were performed according to ISO 14577 under a load of 100 mN. In order to produce more statistically relevant data, the KLA-tencor Express Test software module was employed to enable very large datasets to be generated. Vicker's hardness measurements were performed with a Vickers' diamond indenter at 200 kgf/mm$^2$ with a hold time of 15 seconds. The indentations were examined for conformation with the ASTM C1327. The indentations averaged 15–20 μm in width during the testing. Thirty measurements were performed at different locations of the specimen; the mean and standard deviation are reported. The Vickers indentation test was also carried out following the ASTM standard for measuring the microhardness.

Thermal conductivities were measured using time-domain thermoreflectance [46]. A thin Al transducer (84 ± 4 nm) is thermally evaporated onto the sample. Using a Ti:Sapphire laser emitting a train of <200 fs pulses at a central wavelength of 800 nm and a repetition rate of 80 MHz, the output is divided into a pump and probe path. The pump is modulated at 8.4 MHz to heat the sample, while the probe is used to measure the resulting change in temperature as a function of delay time out to 5.5 ns after pump absorption. The pump and probe $1/e^2$ diameters are 15 and 9 μm, respectively. The volumetric heat capacity was taken to be 2.5 ± 3 J cm$^{-3}$ K$^{-1}$ based on the rule of mixtures average of constituent heat capacities [47].

## 3. Results and Discussion

The XRD pattern shown in Fig. 2 suggests that the $(Mo_{0.2}Nb_{0.2}Ta_{0.2}Ti_{0.2}W_{0.2})Si_2$ specimen made by SPS possesses a hexagonal structure with the space group $P6_222$, or the $CrSi_2$ prototype structure. All peaks, except for one very minor peak, in the XRD pattern (Fig. 2) can be indexed to the hexagonal C40 structure with the ABC stacking sequence, as schematically illustrated in Fig. 1. SEM and EDX maps (Fig. 3) further demonstrated that this five-cation $(Mo_{0.2}Nb_{0.2}Ta_{0.2}Ti_{0.2}W_{0.2})Si_2$ specimen indeed formed a homogenous high-entropy solid solution. This hexagonal C40 structure was further confirmed by EBSD of a polished sample surface (Fig. 4). Lattice parameters of this $(Mo_{0.2}Nb_{0.2}Ta_{0.2}Ti_{0.2}W_{0.2})Si_2$ specimen were determined from the XRD to be: $a = 4.711$ Å and $c = 6.522$ Å.

The formation of a hexagonal C40 crystal structure (with the ABC stacking sequence, as shown in Fig. 1) for this high-entropy $(Mo_{0.2}Nb_{0.2}Ta_{0.2}Ti_{0.2}W_{0.2})Si_2$ specimen is noteworthy and interesting since only two of the five constituent disilicides, NbSi$_2$ and TaSi$_2$ [44,48], form this hexagonal structure at high temperatures. TiSi$_2$ possesses an orthorhombic structure (with the ABCD stacking sequence) [49]. Both MoSi$_2$ and WSi$_2$ normally form tetragonal structures (with the AB stacking sequence), though the hexagonal phases were observed at lower temperatures (below 900˚C and 550˚C, respectively) in thin films [48,50].

This (Mo$_{0.2}$Nb$_{0.2}$Ta$_{0.2}$Ti$_{0.2}$W$_{0.2}$)Si$_2$ represents a new high-entropy ceramic made, with a new, and perhaps the lowest, symmetry among all high-entropy metals and ceramics reported. To date, all except for two high-entropy metals and ceramics reported have cubic symmetries (of simple FCC and BCC [1-8], rocksalt [9,20-29,37,38], fluorite [34,35], pervoskite [30-32], and spinel [33] structures). The two other classes of non-cubic high-entropy materials reported are the metallic HCP HEAs (with the point group of P6$_3$/mmc) [8] and high-entropy metal diborides (with the point group of P6/mmm) [10]. This high-entropy (Mo$_{0.2}$Nb$_{0.2}$Ta$_{0.2}$Ti$_{0.2}$W$_{0.2}$)Si$_2$ has a lower symmetry of P6$_2$22, with a more complex ABC stacking sequence (Fig. 1).

It should be noted that a secondary TiO phase is also present, producing a minor XRD peak as indicated in Fig. 2. We assume that TiO formed because TiSi$_2$ possesses a melting point of ~1500 ˚C [26,27], below our SPS temperature; thus, it is likely that TiSi$_2$ promoted the formation a (transient) liquid phase that assisted sintering but captured surface oxides. TiSi$_2$ has been utilized for liquid assisted sintering of diborides in a similar manner [51,52]. The secondary oxide phases seen in the SEM image (the dark phase in the first panel of SEM image in Fig. 3) are likely SiO$_2$-based glass, which did not show up in XRD (since the amount of TiO identified by XRD, as shown in Fig. 2, is small). ImageJ analysis of the SEM image was performed to estimate the high-entropy silicide phase to be approximately 89 vol. %.

EBSD was utilized to measure the grain size and examine the texture of the sintered (Mo$_{0.2}$Nb$_{0.2}$Ta$_{0.2}$Ti$_{0.2}$W$_{0.2}$)Si$_2$ specimen. An average grain size of 5.4 ± 3.3 µm was found from a measurement of over 5000 grains. No significant texturing was evident in the sample. Two EBSD maps at low and high magnifications, an inverse pole figure, and the measured grain size distribution are shown in Fig. 4.

Nanoindentation hardness measurements of this (hexagonal) high-entropy (Mo$_{0.2}$Nb$_{0.2}$Ta$_{0.2}$Ti$_{0.2}$W$_{0.2}$)Si$_2$ following the ISO 14577 standard using a load of 100 mN produced a value of 16.7 ± 1.9 GPa with a large number of indents. It also measured an elastic modulus of 421 ± 19 GPa, in agreement with the measurements taken by Nakamura et al. for $MoSi_2$ and $WSi_2$ [53]. Moreover, we have conducted Vickers indentation test and measured a microhardness value of 11.6 ± 0.5 GPa from our high-entropy $(Mo_{0.2}Nb_{0.2}Ta_{0.2}Ti_{0.2}W_{0.2})Si_2$ specimen. These measured hardness values are comparable to those reported for $MoSi_2$ in literature, with Newman et al. reporting up to 17.5 ± 2.0 GPa in nanoindentation and Vickers hardness in other prior studies varying from approximately 9 to 14 GPa [53-57]. The microhardness value of our high-entropy $(Mo_{0.2}Nb_{0.2}Ta_{0.2}Ti_{0.2}W_{0.2})Si_2$ specimen is compared with five individual constituent metal disilicides in Table 1. Notably, the Vickers hardness of this high-entropy $(Mo_{0.2}Nb_{0.2}Ta_{0.2}Ti_{0.2}W_{0.2})Si_2$ specimen is higher than the average of the microhardness values of the five individual metal disilicides reported in the literature (which was calculated to be 9.32 GPa by taking and median value for $MoSi_2$).

A significantly reduced thermal conductivity was measured for this (hexagonal) high-entropy $(Mo_{0.2}Nb_{0.2}Ta_{0.2}Ti_{0.2}W_{0.2})Si_2$, in comparison with other metal disilicides [58,59]. Fitting a multilayer heat diffusion model to experimental ratio data [60], the best-fit thermal conductivity was determined to be 6.9 ± 1.1 W $m^{-1}$ $K^{-1}$. In comparison, the thermal conductivity of the (tetragonal) $MoSi_2$ has been measured to be 65 W $m^{-1}$ $K^{-1}$ [58]. The thermal conductivities of (hexagonal) $NbSi_2$, (hexagonal) $TaSi_2$, (orthorhombic) $TiSi_2$, and (tetragonal) $WSi_2$, respectively were measured by Neshpor [59] to be 19.1 W $m^{-1}$ $K^{-1}$, 21.9 W $m^{-1}$ $K^{-1}$, 45.9 W $m^{-1}$ $K^{-1}$, 46.6 W $m^{-1}$ $K^{-1}$, respectively; these reported values from literature are listed in Table 1 to be compared with our measured thermal conductivity of the high-entropy $(Mo_{0.2}Nb_{0.2}Ta_{0.2}Ti_{0.2}W_{0.2})Si_2$. While it is possible that the presence of oxide contamination and porosity reduces the thermal conductivity of our specimen, the measured value of 6.9 ± 1.1 W $m^{-1}$ $K^{-1}$ is significantly lower than reported values of any of the five constituent disilicides. Noting that $NbSi_2$ and $TaSi_2$, which have the same hexagonal crystal and lowest thermal conductivities of 19.1 W $m^{-1}$ $K^{-1}$ and 21.9 W $m^{-1}$ $K^{-1}$, respectively [59], among the five individual disilicides, are perhaps the best benchmarks for comparison. Still, the measured thermal conductivity of this high-entropy $(Mo_{0.2}Nb_{0.2}Ta_{0.2}Ti_{0.2}W_{0.2})Si_2$ is substantial lower (~1/3), presumably due to the high phonon scattering from the five different cations with different masses and a highly distorted lattice. A prior modeling study has demonstrated that >10X reduction in thermal conductivity can be achieved in high-entropy ceramics [13], and similar levels of thermal conductivity reduction was indeed observed in entropy-stabilized oxides [12].

We also attempted to fabricate a $(Mo_{0.2}Nb_{0.2}Ta_{0.2}W_{0.2}Zr_{0.2})Si_2$ specimen via the same procedure, but it did not form a single high-entropy phase. The measured XRD pattern and EDX elemental maps of this $(Mo_{0.2}Nb_{0.2}Ta_{0.2}W_{0.2}Zr_{0.2})Si_2$ specimen are shown in Fig. 5. While a primary hexagonal C40 phase did form, additional Ta-Zr-Si and Nb-Zr-Si rich secondary phases were observed.

## 4. Conclusions

A high-entropy metal disilicide, $(Mo_{0.2}Nb_{0.2}Ta_{0.2}Ti_{0.2}W_{0.2})Si_2$, was successfully synthesized. It possesses a hexagonal structure with a point group of $P6_222$, representing a new high-entropy material family (a high-entropy silicide) and a new non-cubic high-entropy crystal structure made. Characterization by XRD, EDX, and EBSD confirm the presence of a single high-entropy solid-solution phase, albeit some oxide contaminations.

This high-entropy $(Mo_{0.2}Nb_{0.2}Ta_{0.2}Ti_{0.2}W_{0.2})Si_2$ exhibits high nanohardness of $16.7 \pm 1.9$ GPa and Vickers hardness of $11.6 \pm 0.5$ GPa. The measured thermal conductivity of $(Mo_{0.2}Nb_{0.2}Ta_{0.2}Ti_{0.2}W_{0.2})Si_2$ is $6.9 \pm 1.1$ W m$^{-1}$ K$^{-1}$, which is approximately one order of magnitude lower than that of the widely-used tetragonal $MoSi_2$ [58] and ~1/3 of those reported for the hexagonal $NbSi_2$ and $TaSi_2$ with the same crystal structure [59]. The significant reduction in the thermal conductivity can be explained from the high phonon scattering in the high-entropy ceramic.

We acknowledge the partial financial support from an Office of Naval Research MURI program (grant no. N00014-15-1-2863; Program Mangers: Dr. Kenny Lipkowitz and Dr. Eric Wuchina).

**Table 1.** Comparison of the properties of the high-entropy $(Mo_{0.2}Nb_{0.2}Ta_{0.2}Ti_{0.2}W_{0.2})Si_2$ with five individual constituent metal disilicides.

| Compound | Crystal Structure | Vickers Hardness (GPa) | Young's Modulus (GPa) | Thermal Conductivity [W m$^{-1}$ K$^{-1}$] | References |
|---|---|---|---|---|---|
| $MoSi_2$ | Tetragonal | 9-14 | 414 | 65 | [53,54,57] |
| $NbSi_2$ | Hexagonal | 5.4 | 330 | 19.1 | [59,61,62] |
| $TaSi_2$ | Hexagonal | 13 | 338 | 21.9 | [59,62,63] |
| $TiSi_2$ | Orthorhombic | 8.5 | 256 | 45.9 | [59,62,64] |
| $WSi_2$ | Tetragonal | 8.2 | 438 | 46.6 | [53,59,62,65] |
| Rule-of-mixture average of five metal disilicides | | 9.32 | 355 | 40 | |
| $(Mo_{0.2}Nb_{0.2}Ta_{0.2}Ti_{0.2}W_{0.2})Si_2$ | Hexagonal | 11.6 ± 0.5 | 421 ± 19 | 6.9 ± 1.1 | This Study |

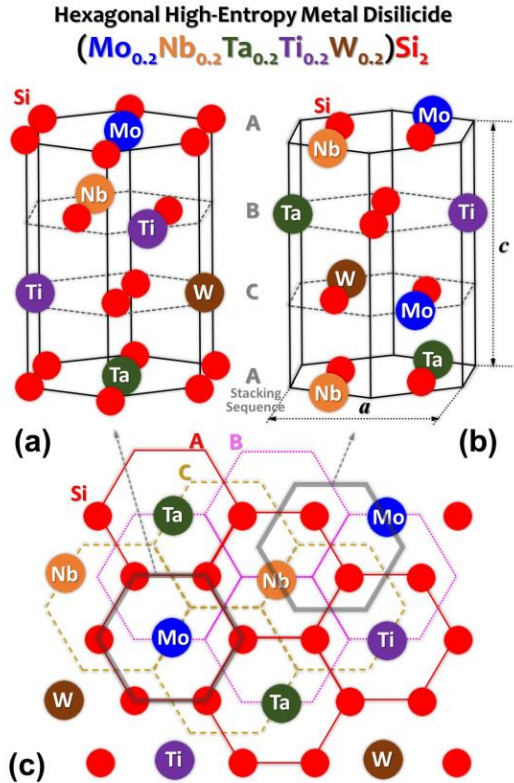

**Figure 1.** Schematic illustration of the atomic structure of the hexagonal high-entropy disilicide with the ABC stacking sequences (*i.e.* the CrSi$_2$ prototype structure). Here, **(a)** and **(b)** are two alternative views of hexagonal cells (but not the unit cells) and **(c)** is an in-plane view, where the positions of both Si and metal atoms are shown for layer A while only the hexagonal Si nets are shown for layers B and C for clarity. The lattice parameters (*a* and *c*) are labeled. Noting that *a* is not the edge of the hexagonal cells shown in (a) and (b), but the distance between two metal cations within the layer.

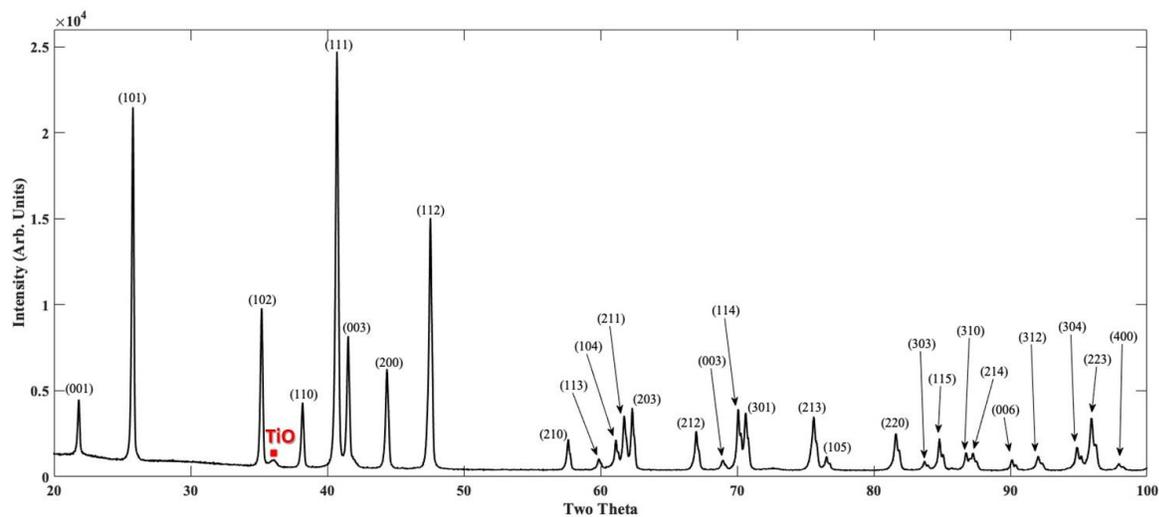

**Figure 2.** XRD pattern of the $(Mo_{0.2}Nb_{0.2}Ta_{0.2}Ti_{0.2}W_{0.2})Si_2$ specimen. Except one minor peak from a secondary hexagonal TiO phase (labeled by the red solid square), all other XRD peaks are indexed to a hexagonal C40 structure (or the $CrSi_2$ prototype structure with the $P6_222$ space group and the $D_6$ point group) with the lattice parameters $a = 4.711$ Å and $c = 6.522$ Å.

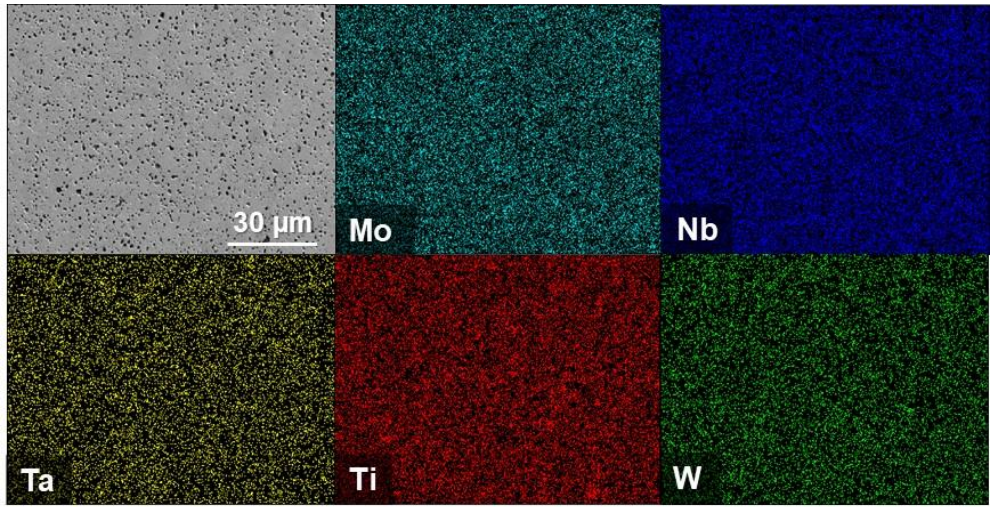

**Figure 3.** SEM micrograph and the corresponding EDX elemental maps of the (Mo$_{0.2}$Nb$_{0.2}$Ta$_{0.2}$Ti$_{0.2}$W$_{0.2}$)Si$_2$ specimen.

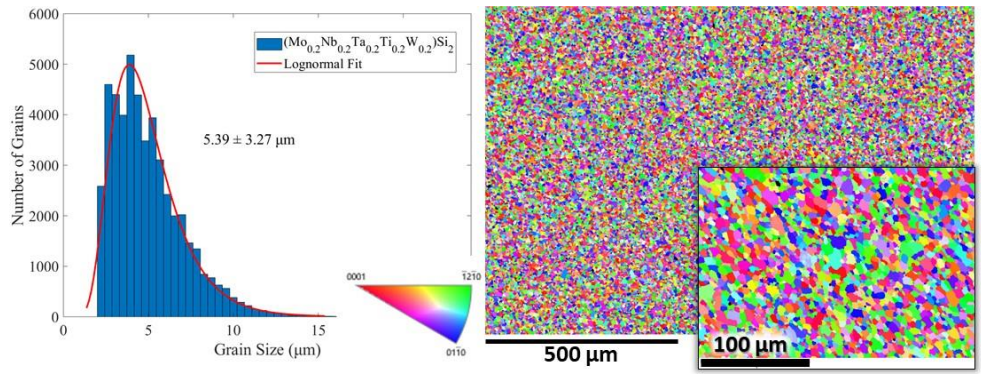

**Figure 4.** EBSD map of >1 mm$^2$ area of the high-entropy (Mo$_{0.2}$Nb$_{0.2}$Ta$_{0.2}$Ti$_{0.2}$W$_{0.2}$)Si$_2$ surface, showing a rather uniform microstructure. No significant texture was observed. The measured grain size distribution is given, which fits a lognormal curve. The inset on the right-bottom corner is an additional BBSD map taken at a higher magnification.

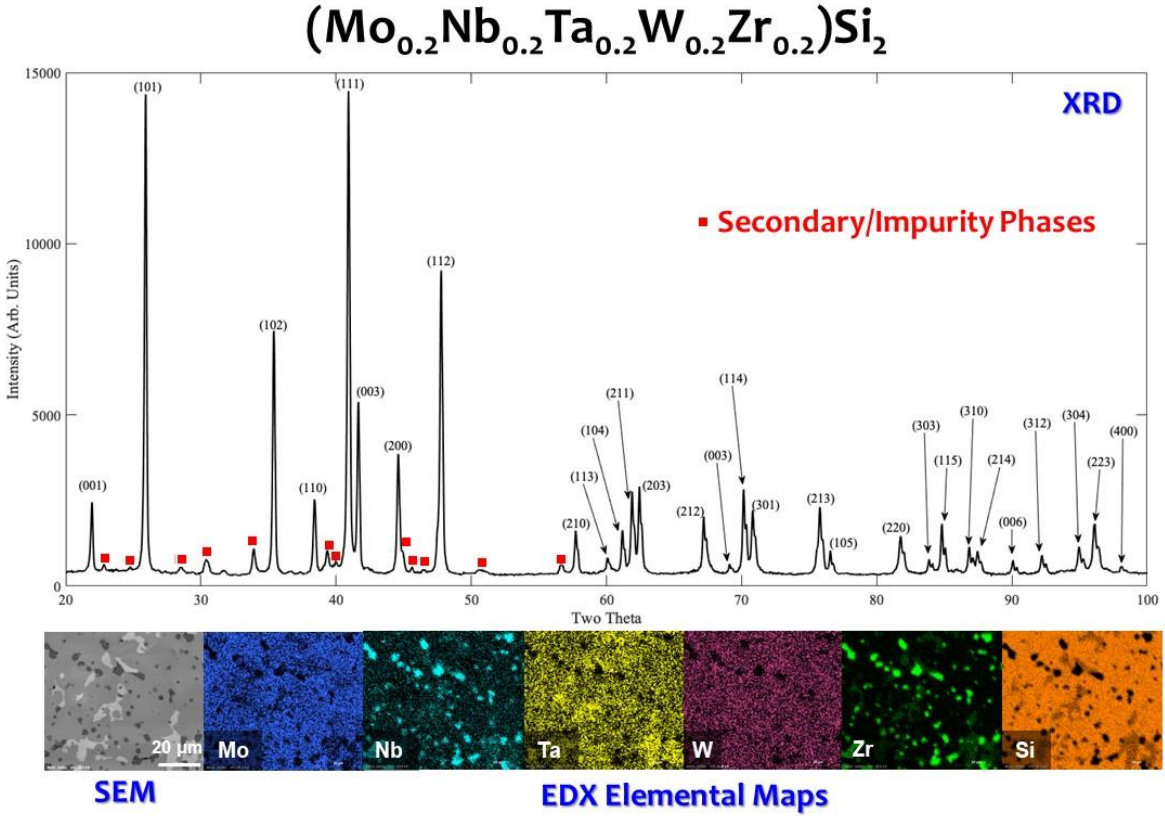

**Figure 5.** XRD pattern, SEM micrograph, and the corresponding EDX elemental maps of the $(Mo_{0.2}Nb_{0.2}Ta_{0.2}W_{0.2}Zr_{0.2})Si_2$ specimen. In addition to a primary hexagonal C40 phase, Ta-Zr-Si and Nb-Zr-Si rich secondary phases were observed.